\documentclass[conference]{IEEEtran}
\IEEEoverridecommandlockouts
\usepackage{cite}
\usepackage{amsmath,amssymb,amsfonts}
\usepackage[table]{xcolor}

\usepackage{amssymb}
\usepackage{graphicx}
\usepackage{wrapfig}
\usepackage{subfig}
\usepackage{url}
\usepackage{verbatim}
\usepackage{multirow}
\usepackage{paralist}
\usepackage{mathptmx}
\usepackage{times}
\usepackage{dirtree}
\usepackage{tabularx}
\usepackage{tikz}
\usepackage[T1]{fontenc}
\usetikzlibrary{arrows,%
                calc,%
                chains,%
                decorations.pathmorphing,%
                decorations.pathreplacing,%
                decorations.shapes,%
                graphs,%
                matrix,%
                positioning,%
                shapes.geometric,%
                shapes,%
                shapes.symbols,%
                trees
   			  }

\usepackage{graphicx}
\usepackage{pgfplots}

\usetikzlibrary{automata,arrows}
\usetikzlibrary{datavisualization}
\usetikzlibrary{datavisualization.formats.functions}
\usetikzlibrary{positioning, calc, backgrounds, arrows}
\usetikzlibrary{shapes.geometric}
\usetikzlibrary{decorations.pathreplacing}
\usetikzlibrary{patterns}

\newtheorem{definition}{Definition}

\newcommand{\optional}[1]{}

\newcommand{\nui}{\bar{u}}
\newcommand{\nuilog}{\bar{\Sigma}}

\newcommand{\reimbursement}{RT}
\newcommand{\studentRecord}{SR}
\newcommand{\scholarshipA}{S1}
\newcommand{\scholarshipB}{S2}

\newcommand{\srx}{SRRT\textsubscript{$+$}}
\newcommand{\sry}{RTSR\textsubscript{$+$}}
\newcommand{\srz}{SRRT\textsubscript{$\parallel$}}
\newcommand{\srk}{RTSR\textsubscript{$\parallel$}}

\usepackage[ruled,vlined,boxed,linesnumbered]{algorithm2e}
\SetKwRepeat{Do}{do}{while}
\SetKwInOut{Input}{input}
\SetKwInOut{Output}{output}

\graphicspath{{images/}}

\begin{document}

\title{Identifying Candidate Routines for Robotic Process Automation From Unsegmented UI Logs}

\author{ \IEEEauthorblockN{	Volodymyr Leno\IEEEauthorrefmark{1}\IEEEauthorrefmark{2},
						Adriano Augusto\IEEEauthorrefmark{1},
						Marlon Dumas\IEEEauthorrefmark{2},
						Marcello La Rosa\IEEEauthorrefmark{1},\\
						Fabrizio Maria Maggi\IEEEauthorrefmark{3}\IEEEauthorrefmark{2},
                        Artem Polyvyanyy\IEEEauthorrefmark{1}}
		\IEEEauthorblockA{				
		\IEEEauthorrefmark{1}University of Melbourne, Australia\\
						\{a.augusto, marcello.larosa, artem.polyvyanyy\}@unimelb.edu.au}
		\IEEEauthorblockA{
		\IEEEauthorrefmark{2}University of Tartu, Estonia\\
						\{leno, marlon.dumas\}@ut.ee}
		%\IEEEauthorblockA{
		%\IEEEauthorrefmark{3}Tecnol{\'o}gico de Monterrey, Mexico\\
		%				luciano.garcia@tec.mx}
		\IEEEauthorblockA{
		\IEEEauthorrefmark{3}Free University of Bozen-Bolzano, Italy\\
						maggi@inf.unibz.it}
}

\maketitle

% !TEX root = main.tex
\begin{abstract}
Robotic Process Automation (RPA) is a technology to develop software bots that automate repetitive sequences of interactions between users and software applications (a.k.a.\ routines). To take full advantage of this technology, organizations need to identify and to scope their routines. This is a challenging endeavor in large organizations, as routines are usually not concentrated in a handful of processes, but rather scattered across the process landscape. Accordingly, the identification of routines from User Interaction (UI) logs has received significant attention. Existing approaches to this problem assume that the UI log is segmented, meaning that it consists of traces of a task that is presupposed to contain one or more routines. However, a UI log usually takes the form of a single unsegmented sequence of events. This paper presents an approach to discover candidate routines from unsegmented UI logs in the presence of noise, i.e.\ events within or between routine instances that do not belong to any routine. The approach is implemented as an open-source tool and evaluated using synthetic and real-life UI logs.
\end{abstract}
%This technology allows organizations to streamline repetitive work in order to focus on the value-adding and knowledge-intensive components of their processes.

% thus reducing the error rates and increasing overall process performance.
%To date, a considerable amount of time is spent on the identification of the candidates for automation.
%The information about such candidates is obtained by interviewing the workers or by observing their work.
%Such approach is not useful, however, in the environments when the number of routines is too large.

\begin{IEEEkeywords}
Robotic process automation, robotic process mining, user interaction log.
\end{IEEEkeywords} 
% !TEX root = main.tex
\section{Introduction}\label{sec:intro}

Robotic Process Automation (RPA) allows organizations to improve their processes by automating repetitive sequences of interactions between a user and one or more software applications (a.k.a.\ routines).
%While there are plenty of enterprise RPA solutions that allow to automate a wide range of routines, they do not allow to identify the candidates for automation.
With this technology, it is possible to automate data entry, data transfer, and verification tasks, particularly when such tasks involve multiple applications.
To exploit this technology, organizations need to identify routines that are prone to automation~\cite{leopold2018identifying}. This can be achieved via interviews or manual observation of workers. In large organizations, however, this approach is not always cost-efficient as routines tend to be scattered across the process landscape. In this setting, manual routine identification efforts can be enhanced via automated methods, for example methods that extract frequent patterns from these User Interaction (UI) logs of working sessions of one or more workers~\cite{lenobise20}.

%which allows to identify the most evident routines, but is not necessarily cost-effective when it comes to identifying

%In addition to being time-consuming, this may also result in the incompleteness of the information about routines to be automated,  which leads to the increased time spent on their automation.
%This work presents an approach to automatically identify candidate routines for automation from the records of the users' interactions with the information systems, also known as user interactions (UI) logs.

%Recent work \cite{lenoDemo2019} proposed to record the execution of a task in the logs of users' interactions with information systems, a.k.a. UI logs.
% Given such logs, routine candidates can be identified automatically.
% For example, one can use sequence mining to extract frequent execution patterns recorded in the log.

Existing methods in this space~\cite{DBLP:conf/iui/DevL17,jimenez2019method,bosco2019,gao2019automated} assume that the event log consists of a set of traces of a task that is presupposed to contain one or more routines. When the log is segmented, the identification of candidate routines boils down to discovering frequent sequential patterns from a collection of sequences, a problem for which a range of algorithms exist.

In practice, though, UI logs are not segmented. Instead, a recording of a working session consists of a single sequence of actions encompassing many instances of one or more routines, interspersed with other events that may not be part of any routine. Traditional approaches to sequential pattern mining, particularly those that are resilient to noise (irrelevant events) are not applicable to such unsegmented logs.

%It captures all the actions performed during one recording session, which may contain multiple executions of a task.
%During the recording, it is not feasible to identify when one task ends, and another starts, and thus, traditional sequence mining approaches can not be applied directly.

This paper addresses this gap by proposing a method to automatically split an unsegmented UI log into a set of segments, each representing a sequence of steps that appears repeatedly in the unsegmented UI log. Once the log is segmented, sequential pattern mining techniques are used to discover candidate routines.
%This approach reduces the time spent on the identification of routines and allows to focus instead on their automation.
The method has been evaluated on synthetic and real-life UI logs in terms of its ability to rediscover routines contained in a log and in terms of scalability.

The paper is structured as follows.  Section \ref{sec:background} provides the necessary background and an overview of related work.  Section \ref{sec:approach} describes the approach, while Section \ref{sec:eval} reports the results of the evaluation. Finally, Section \ref{sec:conclusion} concludes the paper and spells out directions for future work. 
% !TEX root = main.tex
\section{Background and Related Work}\label{sec:background}

%In this section, first, we give an overview of \emph{robotic process mining} and its three major phases~\cite{lenobise20}: i) UI log recording and pre-processing; ii) candidate routine identification; and iii) executable routine discovery. Then, we introduce the problem of segmentation, which is the cornerstone of the \emph{candidate routine identification} phase and the focus of this paper, and we discuss the related work.

%\subsection{Robotic Process Mining}

%This paper addresses the problem of identifying routines in a UI log, which could be potentially be automated using RPA technology. This problem is part

The problem addressed by this paper can be framed in the broader context of Robotic Process Mining (RPM)~\cite{lenobise20}. RPM is a family of methods to discover repetitive routines performed by employees during their daily work, and to turn such routines into software scripts that emulate their execution.
%The routines that can be compiled into software scripts are known as \emph{automatable routines}, and they can be discovered manually or automatically.  Robotic process mining lives within the latter case.
The first step in an RPM pipeline is to record the interactions between one or more workers and one or more software applications~\cite{lenoDemo2019}. The recorded data is represented as a UI log -- a sequence of user interactions (herein called UIs), such as selecting a cell in a spreadheet or editing a text field in a form.
%Even though a UI log may collect UIs from different tasks at the same time, in this paper, we assume that a UI log contains data relating to only one specific task.
The UI log may be filtered to remove irrelevant UIs (e.g. misclicks). Next, it may be decomposed into segments. The discovered segments are scanned to identify routines that occur across these segments. Finally, the resulting routines are analysed to identify those that are automatable and to encode them as RPA scripts. 

%, where (ideally) each segment captures one execution from start to end of the task under observation

%This step is known as \emph{segmentation}.
%Routine variants may be due to different a re-ordering of the UIs as well as to the specialization or generalization of the routine itself.

%DBLP:conf/iui/DevL17,leopold2018identifying,linn2018desktop,jimenez2019method,gao2019automated,

% Leopold et al.~\cite{leopold2018identifying} identify automatable tasks starting from textual process description.

The problem of routine identification from UI logs in the context of RPM has attracted significant attention~\cite{lenoDemo2019}. However, existing approaches for routine identification from UI logs either take as input a segmented UI log~\cite{DBLP:conf/iui/DevL17,linn2018desktop,bosco2019,gao2019automated}, or they assume that the log can be trivially segmented by breaking it down at each point where one among a set of ``start events'' appears~\cite{jimenez2019method}. These start events, which act as delimiters between segments, need to be designated by the user.
%For example, Dev and Liu~\cite{DBLP:conf/iui/DevL17} apply frequent pattern mining to discover routines from segmented UI logs.

Another technique for routine identification~\cite{leopold2018identifying} attempts to identify candidate routines from textual documents -- an approach that is suitable for earlier stages of routine identification and could be used to determine which processes or tasks could be recorded and analyzed in order to identify routines.

%other artifacts besides UI logs,

%The following sentence is useless as it does not explain how the method works or how it relates to the gap we want to fill
%Gao et al.~\cite{gao2019automated} designed and implemented a method for automating repetitive form-filling routines based on \emph{if-then} rules; when a certain condition is observed during the execution of a task, a specific action is resolved, i.e. forms are filled with data. The latter study is also the only one providing a working prototype of their method.

%Leopold et al.~\cite{leopold2018identifying}

Below, we review the literature related to UI log segmentation and identification of routines from (segmented) logs.

\subsection{UI Log Segmentation}

Given a UI log, i.e. a sequence of UIs, segmentation consists in identifying non-overlapping subsequences of UIs, namely \emph{segments}, such that each subsequence represents the execution of a task performed by an employee from start to end. In other words, segmentation searches for repetitive patterns in the UI log. In an ideal scenario, we would observe only one unique pattern (the task execution) repeated a finite number of times. However, in reality, this scenario is unlikely to materialise. Instead, it is reasonable to assume that an employee performing X-times the same task would do some mistakes or introduce variance in how the task is performed.

%han1998mining,
%srivastava2000web,

The problem of segmentation is similar to periodic pattern mining on time series. While several studies addressed the latter problem over the past decades~\cite{cao2007discovery,zhu2017matrix}, most of them require information regarding the length of the pattern to discover, or assume a natural period to be available (e.g. hour, day, week). This makes the adaptation of such techniques to solve the problem of segmentation challenging, unless periodicity and pattern length are known a-priori.

Under the same class of problems, we find web session reconstruction~\cite{spiliopoulou2003framework}, whose goal is to identify the beginning and the end of web navigation sessions in server log data, e.g. streams of clicks and web pages navigation~\cite{spiliopoulou2003framework}. 
%While in our context, segmentation aims at associating each UI in the log to a specific task execution,
%Session reconstruction is a pre-processing step applied to server log data prior to applying web usage mining methods, such as user profiling, statistical analysis, or frequent sequential patterns. 
Methods for session reconstruction are usually based on heuristics that rely either on IP addresses or on time intervals between events The former approach is not applicable in our context (routines in UI logs cannot be segmented based on IP addresses) while the latter approach assumes that users make breaks in-between two consecutive segments -- in our case, two routine instances.

% that users do not make breaks in the middle of a session (in our case, in the middle of a routine) and that they break when switching from one routine instance to another. 

%Even though these methods work only on web browsing data, it is possible to adapt the methods based on time interval heuristics to solve the problem of segmentation, as we show in the evaluation (Section~\ref{sec:eval}).

Lastly, segmentation also relates to the problem of correlation of event logs for process mining. In such logs, each event should normally include an identifier of a process instance (case identifier), a timestamp, an activity label, and possibly other attributes. When the events in an event log do not contain explicit case identifiers, they are said to be uncorrelated.  Various methods have been proposed to extract correlated event logs from uncorrelated ones. However, existing methods in this field either assume that a process model is given as input~\cite{DBLP:conf/caise/BayomieAE16} or that the underlying process is acyclic~\cite{DBLP:conf/bpm/FerreiraG09}. Both of these assumptions are unrealistic in our setting: a process model is not available since we are precisely trying to identify the routines in the log, and a routine may contain repetition. 
%On the other hand, a different assumption can be made in our setting: that the routine instances do not overlap, meaning that the worker performs one routine instance after another.

%In this case, given an event log (i.e. recordings of business process executions), the goal is to associate each event to a specific process execution. BayomieDLM19
%While these assumptions are too strict for our context, we can lay out different reasonable assumptions: i) the UI log records only one task; and ii) different executions of the same task are non-overlapping. In the remainder of this paper, we retain such assumptions.

\subsection{Routine Identification}

%Routines are frequent sequences of actions observed in different segments of the UI log.
%In the ideal case, each segment  of entire task executions (start-to-end) having no variation or noise, entire segments would be identified as routines themselves.
%In general, though, routines are most likely sub-parts of a task execution that are always executed in the same manner, and therefore candidate for automation.

%alves2010gene
%kum2003approxmap,lee2004efficient,ji2007efficient,agrawal1995mining,
%~\cite{wang2004bide,ohlebusch2015alphabet,fumarola2016clofast}

Once the UI log is segmented, the segments are scanned to identify routines. Dev and Liu~\cite{DBLP:conf/iui/DevL17} have noted that the problem of routine identification from (segmented) UI logs can be mapped to that of frequent pattern mining, a well-known problem in the field of data mining~\cite{han2007frequent}. 
%While the former and the latter aim to identifying patterns in different transactions and genes (respectively),
Indeed, the goal of routine identification is to identify repetitive (frequent) sequences of interactions, which can be represented as symbols. In the literature, several algorithms are available to mine frequent patterns from sequences of symbols.
Depending on their output, we can distinguish two types of frequent pattern mining algorithms: those that discover only exact patterns~\cite{lee2004efficient,ohlebusch2015alphabet} (hence vulnerable to noise) and those that allow frequent patterns to have gaps within the sequence of symbols~\cite{wang2004bide,fumarola2016clofast} (hence noise-resilient).
Depending on their input, we can distinguish between algorithms that operate on a collection of sequences of symbols and those that discover frequent patterns from a single long sequence of symbols~\cite{ohlebusch2015alphabet}. The former algorithms can be applied to segmented UI logs while the latter can be applied directly to unsegmented ones. However, techniques that identify patterns from a single sequence of symbols only scale up when identifying exact patterns.

%In Section~\ref{sec:approach}, we discuss the adaptation of two of the existing frequent pattern mining algorithms to solve the problem of routine identification, specifically: i) CloFAST~\cite{fumarola2016clofast}, which discovers frequent patterns over a collection of sequences; and ii) the algorithm of Ohlebusch and Beller~\cite{ohlebusch2015alphabet}, which discovers supermaximal exact patterns from a single sequence of symbols.

%Frequent pattern mining algorithms find home also in the field of web usage mining~\cite{srivastava2000web}, whose goal is to discover useful information from data generated from users web navigation, i.e. server logs.  A specific sub-area of web usage mining focuses on identifying frequent sequential patterns within user sessions, such as frequent sequences of clicks in specific areas of a webpage, or frequent sequences of web pages. However, techniques belonging to this area of web usage mining cannot be used for segmentation, since they leverage web browsing characteristics, while employees typically perform tasks across multiple applications (e.g. web browser and worksheets).

%Such information is then used to study and improve the user web experience.

\section{Approach}\label{sec:approach}

\begin{figure*}[htb]
\centering
\includegraphics[scale=0.6]{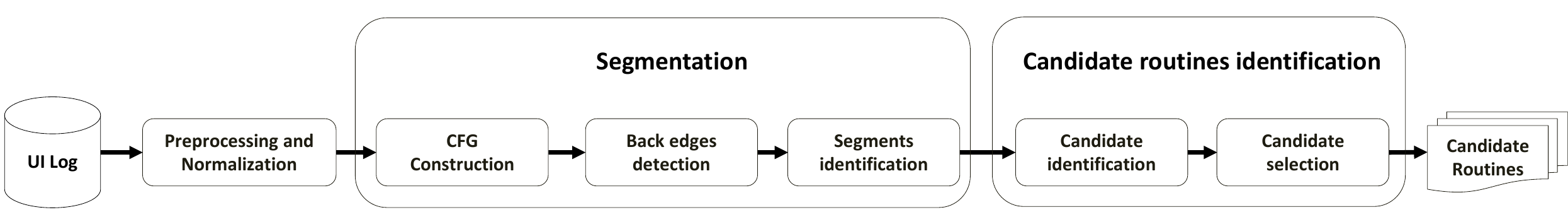}
\caption{Outline of the proposed approach}
\label{fig:approach}
\end{figure*}

In this section,
we describe our approach for identifying candidate routines in UI logs.
As input, the approach takes a preprocessed and normalized UI log and outputs a set of candidate routines.
The approach follows the RPM pipeline~\cite{lenobise20}, and consists of two macro steps.
First, the normalized UI log is decomposed into segments.
Then, candidate routines are identified by mining frequent sequential patterns from the segments.
In this paper, we refer to these two macro steps as \textit{segmentation} and \textit{candidate routines identification}, respectively.
The approach is summarized in \figurename~\ref{fig:approach}.
Next, we describe the steps in details, including the required UI log preprocessing and normalization.

%In this section, we describe two approaches we designed to solve the problem of candidate routines identification from UI logs.
%Both approaches receive in input a preprocessed and normalized UI log and output a set of candidate routines.
%The first approach follows the classic RPM pipeline~\cite{lenobise20}, it identifies the segments in the input UI log and then it scans the segments to detect frequent patterns, i.e. the candidate routines.
%The second approach, instead, does not segment the UI log, but it scans straightforward the latter to identify the candidate routines.
%The outline of these two approaches is captured in \figurename~\ref{fig:approach}.
%In the following, we describe in details each step of the two approaches, starting with the shared ones: UI log preprocessing and normalisation.
%The preprocessing applied to the UI log is an adaptation of~\cite{leno2020iaaa}, which we describe in the next subsection.
%We describe preprocessing step in Section \ref{sec:preprocessing}, and log parsing step in Section \ref{sec:eventsParsing}. Then, in Section \ref{sec:approach1} and Section \ref{sec:approach2} we present two outlined approaches for candidate routines identification.

%\begin{figure*}[htb]
%	\centering
%	\subfloat[Approach 1: Segmentation + Sequence mining]{
%		\includegraphics[scale=0.7]{images/Approach1.pdf}\label{fig:approach1}
%	}
%	\hspace{3mm}
%	\subfloat[Approach 2: Repeats mining]{
%		\includegraphics[scale=0.7]{images/Approach2.pdf}\label{fig:approach2}
%	}
%\caption{Outline of the proposed approaches}
%\vspace{-4mm}
%\label{fig:approach}
%\end{figure*}

%\begin{figure*}[htb]
%\centering
%\includegraphics[scale = 0.35]{images/Approach1.pdf}
%\caption{Outline of the proposed approaches}
%\label{fig:approach}
%\end{figure*}

\subsection{UI Log Preprocessing and Normalization}
\label{sec:preprocessing}

Before we proceed, we give formal definitions necessary to support subsequent discussions.

\begin{definition}[\textbf{User Interaction (UI)}]
A \emph{user interaction} (UI) is a tuple $u = (t, \tau, P, Z, \phi)$, where:
$t$ is a UI timestamp;
$\tau$ is a UI type (e.g. click button, copy cell);
$P_{\tau}$ is a set of UI parameters (e.g. button name, worksheet name, url, etc.);
$Z$ is a set of UI parameters values; and
$\phi : P_{\tau} \rightarrow Z$ is a function that maps UI parameters onto values.
\end{definition}

\begin{definition}[\textbf{UI Log}]
A UI Log $\Sigma$ is a sequence of user interactions $\Sigma = \langle u_1, u_2, \dots, u_n \rangle$, ordered by timestamp,
i.e. $u_{i\mid t} < u_{j\mid t}$ for any $i,j \mid 1 \leq i < j \leq n$. In the reminder of this paper, we refer to UI log also as log.
\end{definition}

Ideally, the UIs recorded in a UI log should capture only the execution of the task under recording.
However, a log often contains also UIs that do not bring any value to the recorded task, and that should not have been executed in the first place.
We can consider such UIs \emph{noise}, some common examples include:
% not relevant for the outcome of the recorded task.
%we differentiate such UIs into \emph{noise} and \emph{waste}. %two main class of UIs belonging to this latter class.
%UIs that are not necessary for the completion of the recorded task includes:
%and are performed intentionally by the user correspond to \textit{noise},
%for example, an employee
an employee replying to incoming emails and/or browsing the web while executing a different task;
or an employee committing mistakes, e.g. filling a text field with an incorrect value, or copying an incorrect text or cell in a spreadsheet.
%UIs that are not necessary for the completion of the recorded task and are performed unintentionally by the user correspond to \textit{waste},
%for example, an employee
%or perform redundant actions (e.g., navigating across cells in a spreadsheet without copying their content).
%Such activities constitute \textit{waste}.
To reduce the impact of noise on routines identification, we preprocess the log.
%Therefore, the UI log requires first step in our overall approach is \textit{preprocessing},
The preprocessing we apply consists in identifying and removing redundant UIs that are clearly overwritten by successive UIs,
such as the case of a double \emph{CTRL+C} performed in sequence on different text fields.
To identify such patterns of UIs, we rely on regular expressions by applying the methods described in~\cite{leno2020iaaa}.
%First, all the events in UI log are ordered by timestamp as they are coming from various unsynchronized applications.
%When the order of events in the log is established,
%To do so, we look for specific values in the UIs payload using regular expressions,
%and we replace such patterns of the type find-and-replace to filter out redundant actions.
%These rules describe redundant patterns and how to deal with them.
%For example, when a copy action is followed by another copy and there is no paste action,
%then the first copy is redundant and must be removed.
%On this stage, only read (e.g. copy a content of a cell) and navigation (e.g. click on a cell) actions are removed.

After the preprocessing, the vast majority of UIs in the log are unique, because they differ by their payload.
Even UIs capturing the same action within the same task execution (or different task executions) appears to be different.
To discover each task execution (from start to end) recorded in the UI log,
we need to detect all the UIs that even having different payload correspond to the same action within the task execution.

%To do so, we need to define the concepts of \emph{UI parameter value diversity}, \emph{UI context parameter}, and \emph{normalized UI}.

%\begin{definition}[\textbf{UI parameter value diversity}]
%Given a UI log $\Sigma = \langle u_1, u_2, \dots, u_n \rangle$,
%for each set of UIs of the same type $U_{\hat{\tau}} = \{(t, \tau, P_{\tau}, Z, \phi) \in \Sigma \mid \tau = \hat{\tau}\}$,
%and for each $p \in P_{\hat{\tau}}$ we define the parameter value diversity of $p$ as $d(p) = \frac{\left| m \right|}{\left| U_{\hat{\tau}} \right|}$,
%where $m$ is the total number of unique values that $p$ assumes over all the UIs of type $\hat{\tau}$.
%\end{definition}

%\begin{definition}[\textbf{UI context parameter}]
%Given a type of UI $\tau$ and the corresponding set of parameters $P_{\tau}$,
%we say that $p \in P_{\tau}$ is a context parameter for any UI of type $\tau$ iff $d(p) < \epsilon$, where $\epsilon \in [0,1]$.
%Given a UI $u = (t, \tau, P_{\tau}, Z, \phi))$, we define its \textit{UI context} as $\bar{P_{\tau}} = \{p \in P_{\tau} \mid d(p) < \epsilon\} $;
%i.e. $\bar{P_{\tau}}$ is the set of all the context parameters of $u$.
%\end{definition}

To do so, we need to introduce the concepts of \emph{UI data parameter}, \emph{UI context parameter}, and \emph{normalized UI}.
Given a UI, its parameters can be divided into two types: \emph{data parameter} and \emph{context parameter}.
The \emph{data parameters} store the data values that are used during the execution of a task, e.g. the value of text fields or copied content.
The \emph{data parameters} usually have different values per task execution.
%By contrast, the context parameters store the information about environment.
%While the UI type describes what action was performed (e.g. click button, copy),
By contrast, the \emph{context parameters} capture where the UI was performed, e.g. the application and the location within the application.
The \emph{context parameters} are likely to have always the same values even in different task executions.
For example, a UI of type \emph{copy} includes the following parameters:
\emph{target-application} (e.g. the browser), \emph{user}, \emph{element id} (e.g. the id of a text field in the browser), \emph{copied content}.
Here, \emph{target-application} and \emph{element id} are the context parameters, while \emph{copied content} is the data parameter.
Naturally, different type of UIs are characterized by different context parameters,
e.g. a UI in a spreadsheet has the following context parameters: \emph{spreadsheet name} and \emph{cell location}.
\footnote{The context parameters are selected by the domain expert.}
%while, a UI in a web-based application would have \emph{URL}, \emph{element id}, name of a web element, hyperlink, etc.
%This resembles automated process discovery, where the business analyst has to select the timestamp, case identifier and activity name attributes.

%We note that certain UI parameters (e.g. \emph{copied content}) are likely to have different values in each execution of the action within different task executions, while other parameters (e.g. \emph{element id}) are likely to have always the same values even within different task executions.
%Given that the latter type of parameters would show a low diversity they would be considered context parameters.
%Following on this example, the context parameters of a UI of type \emph{copy} would include: \emph{target-application}, and \emph{element id}.
%By definition, UI context parameters are determined by the value assigned to $\epsilon$, so that changing its value may result in identifying different context parameters. In Section~\ref{sec:eval}, we discuss what value to assign to $\epsilon$, in order to obtain the optimal results when identifying routines.

\begin{definition}[\textbf{Normalized UI}]
Given a UI $u = (t, \tau, P_{\tau}, Z, \phi)$,
we define its normalization as $\bar{u} = (t, \tau, \bar{P_{\tau}}, \bar{Z}, \phi)$;
where $\bar{Z} \subseteq Z$ contains only the values of the parameters in $\bar{P}$, where $\bar{P}$ is a set of context parameters.
Given two normalized UIs, $u_1 = (t_1, \tau, P_{\tau}, \bar{Z_1}, \phi_1), u_2 = (t_2, \tau, P_{\tau}, \bar{Z_2}, \phi_2)$,
the equality relation $u_1 = u_2$ holds iff $\forall p \in P_{\tau} \Rightarrow \phi_1(p) = \phi_2(p)$.
\end{definition}

Given a UI log $\Sigma = \langle u_1, u_2, \dots, u_n \rangle$, we normalize it by normalizing all the recorded UIs.
The resulting normalized UI log is $\bar{\Sigma} = \langle \bar{u}_1, \bar{u}_2, \dots, \bar{u}_n \rangle$.
Table~\ref{tab:uilog} and~\ref{tab:norm-uilog} show, respectively, a small fragment of a UI log and its normalized version.
Intuitively, in a normalized UI log, the chances that two executions of the same routine have the same sequence (or set) of normalized UIs are high,
because normalized UIs have only context parameters. %,which have low diversity.
In the next steps, we leverage such a characteristic of the normalized UI log to identify its segments (i.e. start and end of each executed task),
and the routine(s) within the segments.

\begingroup
\renewcommand{\arraystretch}{1.6}
\begin{table}[hbtp]
	\centering
	\scalebox{0.65}{
    \begin{tabular}{c|c|c|l|l|l}
    & \textbf{UI}
    & \textbf{UI}
    & \multicolumn{3}{|c}{\textbf{UI Parameters and Values}}
    \\\cline{4-6}

    & \textbf{Timestamp}
    & \textbf{Type}
    & \textbf{\textbf{$P_1$: Application}}
    & \textbf{\textbf{$P_2$: Element Label}}
    & \textbf{\textbf{$P_3$: Element Value}}\\\hline

    1 & 2019-03-03T19:02:18 & Click button & Web & New Record & --\\\hline
    2 &  2019-03-03T19:02:23 & Edit field & Web & Full Name & Albert Rauf \\\hline
    3 & 2019-03-03T19:02:27 & Edit field & Web & Date & 11-04-1986 \\\hline
    4 & 2019-03-03T19:02:39 & Edit field & Web & Phone & + 61 043 512 4834 \\\hline
    5 &  2019-03-03T19:02:47 & Click button & Web & Submit & --\\\hline
    6 & 2019-03-03T19:02:58 & Click button & Web & New Record & --\\\hline
    7 & 2019-03-03T19:03:13 & Edit field & Web & Date & 20-06-1987 \\\hline
    8 & 2019-03-03T19:03:24 & Edit field & Web & Phone & +61 519 790 1066 \\\hline
    9 & 2019-03-03T19:03:43 & Edit field & Web & Full Name & Audrey Backer \\\hline
    10 & 2019-03-03T19:04:10 & Click button & Web & Submit & --\\\hline
    \end{tabular}
    }
  	\caption{Fragment of a UI log.}\label{tab:uilog}
  	\vspace{-2mm}
\end{table}
\endgroup

\begingroup
\renewcommand{\arraystretch}{1.6}
\begin{table}[hbtp]
	\centering
	\scalebox{0.65}{
    \begin{tabular}{c|c|c|l|l}
    & \textbf{UI}
    & \textbf{UI}
    & \multicolumn{2}{|c}{\textbf{UI Parameters and Values}}
    \\\cline{4-5}

    & \textbf{Timestamp}
    & \textbf{Type}
    & \textbf{\textbf{$P_1$: Application}}
    & \textbf{\textbf{$P_2$: Element Label}}\\\hline

    1 & 2019-03-03T19:02:18 & Click button &Web & New Record \\\hline
    2 &  2019-03-03T19:02:23 & Edit field &  Web & Full Name \\\hline
    3 & 2019-03-03T19:02:27 & Edit field & Web & Date \\\hline
    4 & 2019-03-03T19:02:39 & Edit field & Web & Phone \\\hline
    5 &  2019-03-03T19:02:47 & Click button & Web & Submit \\\hline
    6 & 2019-03-03T19:02:58 & Click button & Web & New Record \\\hline
    7 & 2019-03-03T19:03:13 & Edit field & Web & Date \\\hline
    8 & 2019-03-03T19:03:24 & Edit field & Web & Phone \\\hline
    9 & 2019-03-03T19:03:43 & Edit field & Web & Full Name \\\hline
    10 & 2019-03-03T19:04:10 & Click button & Web & Submit \\\hline
    \end{tabular}
    }
  	\caption{Normalised UI log.}\label{tab:norm-uilog}
  	\vspace{-2mm}
\end{table}
\endgroup

\subsection{Segmentation}
\label{sec:segmentation}

%The main idea of this approach is to construct a \textit{control-flow graph (CFG)} from a normalized UI log and then analyse this graph to find repetitive segments, that are represented by loops in the graph. Given a set of segments, frequent sequential patterns are identified and these patterns represent candidate routines. Before we move to description of approach, we provide all the necessary terminology.

Before describing in details the segmentation step, we formally define the necessary cornerstone concepts.
%The segmentation step starts from construction of a \textit{control-flow graph (CFG)} from the normalized UI log.

%\begin{definition}[\textbf{Control-flow graph}]
%A control flow graph G = (V, E, e) is a directed graph, where V denotes a set of vertices, E is a set of edges, and $e \in V$ is an entry vertex.
%\end{definition}

%A node is said to \textit{have predecessors} if it is reachable from any other node.

\begin{definition}[\textbf{Directly-follows relation}]
Let $\bar{\Sigma} = \langle \bar{u}_1, \bar{u}_2, \dots, \bar{u}_n \rangle$ be a normalized UI log.
Given two normalized UIs, $\bar{u}_x, \bar{u}_y \in \bar{\Sigma}$,
we say that $\bar{u}_y$ directly-follows $\bar{u}_x$, i.e. $\bar{u}_x \leadsto \bar{u}_y$,
iff $\bar{u}_{x\mid t} < \bar{u}_{y\mid t} \wedge \nexists \bar{u}_z \in \Sigma \mid \bar{u}_{x\mid t} \geq \bar{u}_{z\mid t} \geq \bar{u}_{y\mid t}$.
%for any $x,y,z \in [1, n]$
\end{definition}

\begin{definition}[\textbf{Control-flow Graph (CFG)}]
Given a normalized UI log, $\bar{\Sigma} = \langle \bar{u_1}, \bar{u_2}, \dots, \bar{u_n} \rangle$,
let $\bar{A}$ be the set of normalized UIs in $\bar{\Sigma}$.
A control-flow graph (CFG) is a tuple $G = (V, E, \hat{v}, \hat{e})$, where:
$V$ is the set of vertices of the graph, each vertex maps one normalized UI in $\bar{A}$;
$E \subseteq V \times V$ is the set of edges of the graph, and each $(v_i, v_j) \in E$ represents a directly-follows relations between the UIs mapped by $v_i$ and $v_j$;
$\hat{v}$ is the graph \emph{entry vertex}, such that $\forall v \in V \nexists (v, \hat{v}) \in E \wedge \nexists (\hat{v}, v) \in E$;
while $\hat{e} = (\hat{v}, v_0)$ is the graph \emph{entry edge}, such that $v_0$ maps $\bar{u_1}$.
We note that $\hat{v} \notin V$, and $\hat{e} \notin E$, since they are artificial elements of the graph.
\end{definition}

\begin{definition}[\textbf{CFG Path}]
Given a CFG $G = (V, E, \hat{v}, \hat{e})$,
a CFG path is a sequence of vertices $p_{v_1,v_k} = \langle v_1, \dots, v_k \rangle$
such that for each $i \in [1,k-1] \Rightarrow v_i \in V \cup \{ \hat{v} \} \wedge \exists (v_i, v_{i+1}) \in E \cup \{ \hat{e}\}$.
\end{definition}

%\begin{definition}[\textbf{Entry vertex}]
%An entry vertex entry(G) in a graph G is a vertex that has no predecessors and for each $v \in V$, there exists a path from entry(G) to v. A node $n \is G(V)$ is said to \textit{have predecessors}
%\end{definition}

%A set X of vertices is said to be \textit{strongly connected} if there exists a path consisting only of vertices in X, between any two vertices of X.
%X is said to be a \textit{strongly connected component} of the graph if X is strongly connected and no proper superset of X is strongly connected.
%A strongly connected component is said to be \textit{non-trivial} if it consists of two or more vertices or if it consists of one vertex that has a self loop (an edge from the vertex to itself). Given a graph G, SCC(G) denote the set of non-trivial strongly connected components of G.

\begin{definition}[\textbf{Strongly Connected Component (SCC)}]
Given a graph $G = (V, E, \hat{v}, \hat{e})$, a strongly connected component (SCC) of G is a pair $\delta = (\bar{V}, \bar{E})$,
where $\bar{V} = \{ v_1, v_2, \dots, v_m \} \subseteq V$ and $\bar{E} = \{ e_1, e_2, \dots, e_k \} \subseteq E$
such that $\forall v_i, v_j \in \bar{V} \exists p_{v_i,v_j} \mid \forall v \in p \Rightarrow v \in \bar{V}$.
Given an SCC $\delta = (\bar{V}, \bar{E})$, we say that $\delta$ is \emph{non-trivial} iff $\left| \bar{V} \right| > 1$.
Given a graph $G$, $\Delta_G$ denotes the set of all the non-trivial SCCs in G.
\end{definition}

%such that each unique normalized UI is a vertex in a graph, and if two UIs are in directly-follows relation there is an edge between the corresponding vertices.
%It may be possible that constructed graph will have no entry vertex.
%For example, in the case when all UIs in the log represent the execution of routine and there are no any other UIs that do not belong to routine,
%the last vertex may be connected to the first one by directly-follows relation.
%To ensure the presence of entry vertex, we add an artificial vertex $\bar{\upsilon}$ and mark it as entry.
%To connect it to the graph, we add a directly follows relation $\bar{\upsilon} \Rightarrow \bar{u_1}$,
%where $\bar{u_1}$ is the first normalized UI in the log.
\begin{figure}[htb]
\centering
\includegraphics[scale = 0.5]{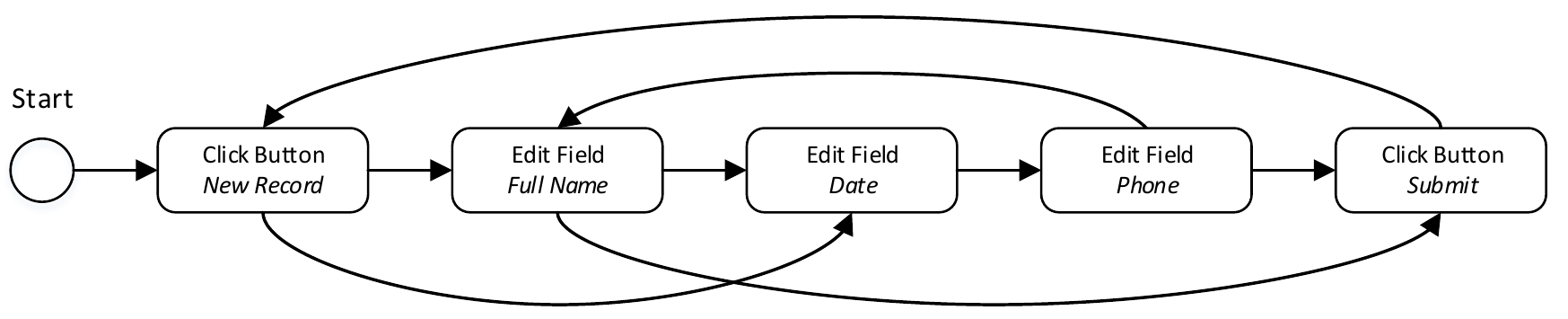}
\caption{Example of a control-flow graph}
\label{fig:cfg}
\end{figure}
%Given preprocessed and normalized UI log we build a control-flow graph such that all unique UIs in their compact representation will represent the set of vertices in the graph, and all directly follows relations between such UIs will constitute the set of edges in the graph. We add artificial entry vertex that is in directly follows relation with the first UI in the log, and construct a dominator tree for constructed control-flow graph.

The segmentation step starts with the construction of the CFG of the input normalized UI log.
%First, we create the artificial entry vertex, the vertex mapping the first normalized UI in the normalized UI log,
%and the artificial entry edge, then, we set to scan the whole normalized UI log.
%For each normalized UI that is not mapped to a vertex in the CFG, we create a new vertex,
%and for each edge between two vertices that does not exist, we create a new edge $(v_1, v_2)$.
%However, if $v_2$ is an existing vertex,
%regardless of whether $(v_1, v_2)$ already in $E$ or not, $(v_1, v_2)$ will be a loop in the CFG.
%which is captured by Algorithm~\ref{alg:buildCFG}.
%First, we create the artificial entry vertex, the vertex mapping the first normalized UI in the normalized UI log,
%and the artificial entry edge (line~\ref{alg:CFGinit1} to~\ref{alg:CFGinit2}), then, we set to scan the whole normalized UI log.
%For each normalized UI that is not mapped to a vertex in the CFG, we create a new vertex (line~\ref{alg:newvertex1} to~\ref{alg:newvertex2}).
%and for each edge between two vertexes that does not exist, we create a new edge $(v_1, v_2)$ (line~\ref{alg:newedge1}).
%However, if $v_2$ is an existing vertex (i.e. $\lambda[\nui]$ is not null, line~\ref{alg:newvertex11}),
%regardless of whether $(v_1, v_2)$ already in $E$ or not, $(v_1, v_2)$ will be a loop in the CFG.
It is common that a CFG contains loops, since a loop represents the start of a new execution of the task recorded in the UI log.
Indeed, in an ideal scenario, once a task execution ends with a certain UI (a vertex in the CFG),
the next UI (i.e. the first UI of the next task execution) should already be mapped in the CFG and a loop will be generated.
In such case, all the vertices contained in the loop represent the corresponding UIs that belong to the task.
If several different tasks are recorded in the UI log, the graph would contain multiple disjoint loops,
while if a task has repetitive subtasks there would be nested loops.
\figurename~\ref{fig:cfg} shows the CFG generated from the normalized UI log captured in Table~\ref{tab:norm-uilog}.

\begin{figure}[htb]
\centering
\includegraphics[scale = 0.55]{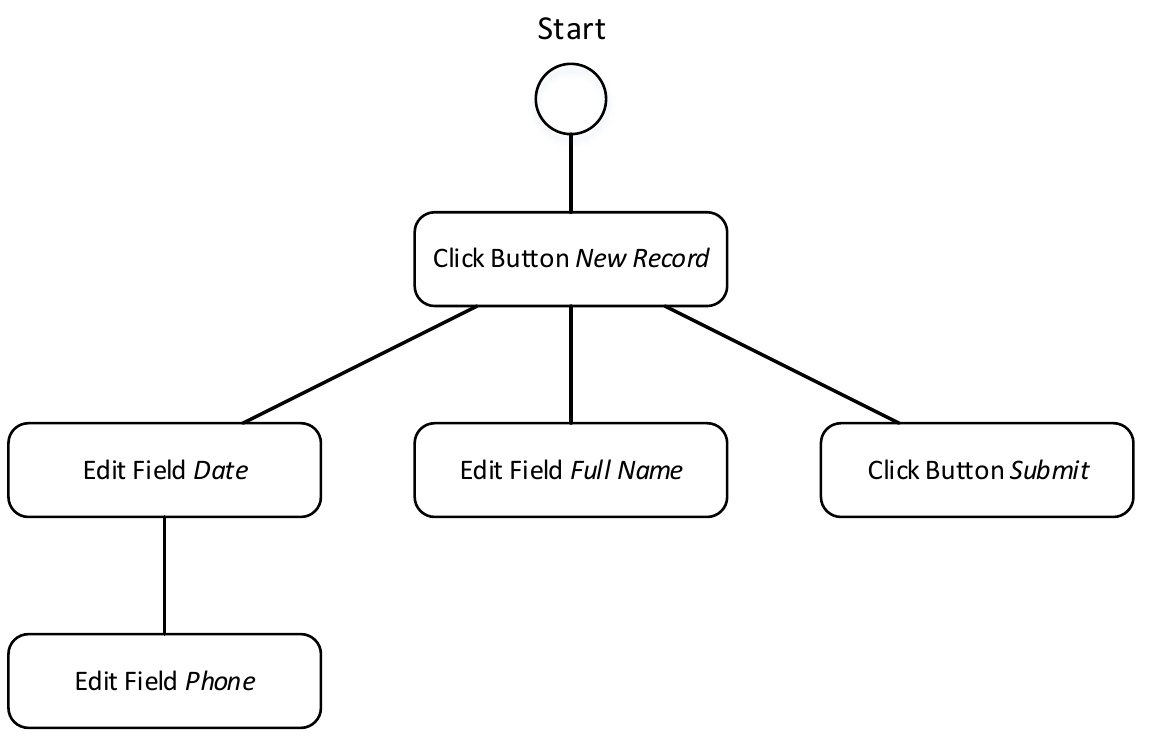}
\caption{Dominator tree}
\label{fig:domTree}
\end{figure}

After the CFG is generated, we turn our attention to the identification of its back-edges,
which can be detected by analysing the SCCs of the CFG, as described in Algorithm~\ref{alg:beDetection} and Algorithm~\ref{alg:analyseSCC}.
Given a CFG $G = (V,E,\hat{v},\hat{e})$, we first build its dominator tree $\Theta$ (Algorithm~\ref{alg:beDetection}, line~\ref{alg:domTree}),
which captures domination relations between the vertices of the CFG.
\figurename~\ref{fig:domTree} shows the dominator tree of the CFG in \figurename~\ref{fig:cfg}.

Then, we discover the set of all non-trivial SCCs ($\Delta_G$) by applying Kosaraju's algorithm \cite{sharir1981strong}
and removing the trivial SCCs (Algorithm~\ref{alg:beDetection}, line~\ref{alg:scc}).
For each $\delta = (\bar{V}, \bar{E}) \in \Delta_G$ we discover its \emph{header} using the dominator tree (Algorithm~\ref{alg:analyseSCC}, line~\ref{alg:header}).
The header of a $\delta$ is a special vertex $\hat{h} \in \bar{V}$,
such that $\forall p_{\hat{v},v} \mid v \in \bar{V} \Rightarrow \hat{h} \in p_{\hat{v},v}$,
i.e. the \emph{header} $\hat{h}$ is the SCC vertex that dominates all the other SCC vertices.
Once we have $\hat{h}$, we can identify the back-edges as $(v,\hat{h})$ with $v \in \bar{V}$ (line~\ref{alg:incoming}).
Finally, the identified back-edges are stored and removed (see line~\ref{alg:backEdges} and ~\ref{alg:edgesSub})
in order to look for nested SCCs and their back-edges by recursively executing Algorithm~\ref{alg:analyseSCC} (line~\ref{alg:recursion}),
until no more SCCs and back-edges are found.
However, if we detect an SCC that does not have a header vertex (i.e. the SCC is irreducible), we cannot identify the SCC back-edges.
In such a case, we collect via a depth-first search of the CFG the edges $(v_x, v_y) \in \bar{E}$ such that $v_y$ is topologically deeper than $v_x$, we call these edges \emph{loop-edges} of the SCC (line~\ref{alg:loops}).
Then, out of all the loop-edges, we store (and remove from the SCC) the one having target and source connected by the longest \emph{simple path} entirely contained within the SCC (see line~\ref{alg:deepestEdge} to ~\ref{alg:removeEdge}).
%we simply remove an edge whose source is topologically deeper than its target in order to investigate possible inner SCCs.
%All such edges in SCC are identified by applying depth-first search of the CFG and
%the one that covers the largest amount of vertices in its \emph{longest simple path} between target and source vertices is selected and removed from SCC (Algorithm~\ref{alg:analyseSCC}, line 7 to 9).

%\begin{definition}[\textbf{Longest simple path}]
%Given a CFG $G = (V, E, \hat{v}, \hat{e})$,
%we define the longest simple path as a sequence of vertices $p_{x,y} = \langle x, v_1, \dots, v_m, y \rangle$
%such that $v_i \neq v_j$ for any $i, j \in [1, m]$ and there is no path $l_{x,y} = \langle x, v_1, \dots, v_n, y \rangle$, where $n > m$.
%\end{definition}

Given the CFG presented in \figurename~\ref{fig:cfg} and its corresponding dominator tree (see \figurename~\ref{fig:domTree}),
we immediately identify the SCC that consists of all the vertices except the \emph{entry vertex}.
Then, by applying Algorithm~\ref{alg:analyseSCC}, we identify:
the SCC header -- \emph{Click Button [New Record]}; and
the only back-edge -- (\emph{Click Button [Submit]}, \emph{Click Button [New Record]}), which we save and remove from the SCC.
After the removal of this back-edge, we identify the nested SCC that contains all the three \emph{Edit Field} UIs.
This second SCC, however, does not have a header because it is irreducible,
due to its multiple entries (\emph{Edit Field [Full Name]} and \emph{Edit Field [Date]}).
However, by applying the depth-first search, we identify as candidate loop-edge for removal: (\emph{Edit Field [Phone]}, \emph{Edit Field [Full Name]}).
After we remove this edge from the CFG, no SCCs are left so that Algorithm~\ref{alg:analyseSCC} stops its recursion.

%If SCC has only one entry point then all the corresponding incoming edges are considered to be \textit{back-edges}.
%We save all identified back-edges and then remove them to break SCC.
%If SCC has multiple entry points it means that there are no back-edges.
%However, we still have to remove the edge that creates a cycle in order to break SCC.
%All such edges in SCC are identified by applying DFS traversal and the one that covers the largest amount of vertices in its \textit{longest simple path} between target and source vertices is selected and removed from SCC.

%\begin{definition}[\textbf{Longest simple path}]
%Longest simple path from vertex x to vertex y in graph G is a path $p = \langle x, p_1, \dots, p_m, y \rangle$ such that $p_i \neq p_j$ where $i \neq j$ for any $i,j \in [1, m]$ and there is no path $l = \langle x, p_1, \dots, p_n, y \rangle$ where $n > m$.
%\end{definition}
%
%Next we identify strongly connected components in SCC without removed edges and perform all the procedure described earlier until there will be only trivial SCCs left. %In such way we can obtain the hierarchy of back-edges that can be used to identify routines and their subroutines.

\begin{algorithm}[btp]
{   \scriptsize
	\Input{CFG $G$}
	\Output{Back-edges Set $B$}
	\BlankLine
	
	$B \leftarrow \varnothing$\;
	Dominator Tree $\Theta \leftarrow$ computeDominatorTree($G$)\;\label{alg:domTree}
	Set $\Delta_G \leftarrow$ findSCCs($G$)\; \label{alg:scc}
	
	\lForEach{$\delta \in \Delta_G$}{
		AnalyseSCC($\delta$, $\Theta$, $B$)\; \label{alg:analyse}
	}
	\BlankLine
	\Return $B$\;
	
	\caption{Detect Back-edges}\label{alg:beDetection}
}
\end{algorithm}

\begin{algorithm}[btp]
{
    \scriptsize
	\Input{SCC $\delta = (\bar{V}, \bar{E})$, Dominator Tree $\Theta$, Back-edges Set $B$}
	\BlankLine

		Header $\hat{h} \leftarrow$ findHeader($\delta$, $\Theta$)\; \label{alg:header}
		\eIf{$\hat{h} \neq$ null}{
			Set $I \leftarrow$ getIncomingEdges($\delta$, $\hat{h}$)\; \label{alg:incoming}
			$B \leftarrow B \cup I$\;  \label{alg:backEdges}
			$\bar{E} \leftarrow \bar{E} \setminus I$\; \label{alg:edgesSub}
		}{
			Set $L \leftarrow$ findLoopEdges($\delta$)\; \label{alg:loops}
			Edge $e \leftarrow$ getTheDeepestEdge($\delta$, $L$)\; \label{alg:deepestEdge}
			remove $e$ from $\bar{E}$\; \label{alg:removeEdge}
		}
		Set $\Delta_\delta \leftarrow$ findSCCs($\delta$)\; \label{alg:findSCCs}
		\lForEach{$\gamma \in \Delta_\delta$}{
			AnalyseSCC($\gamma$, $\Theta$, $B$)\; \label{alg:recursion}
		}
	
	\caption{Analyse SCC}\label{alg:analyseSCC}
}
\end{algorithm} 

%Given a list of back-edges, the next step is to identify possible starts and ends of segments and label UIs in the log correspondingly.
%For back-edge $b = (x, y)$, y is the vertex that denotes possible start of a segment and x is the possible end of a segment.
%Given a set of starts S and ends E, UI u is labeled as start (``S") if $u|\tau + u|c \in S$.
%If $u|\tau + u|c \in E$ then UI u is labeled as end (``E").
%Otherwise, it is labeled as intermediate UI (``-").

%The next step is to identify all the segments in UI log.
%We scan all UIs starting from the beginning of the log and if UI is labeled as start and there is no ongoing segment we ``open" a new segment.
%If we find a UI labeled as end, then we check whether there is exist a back-edge $b = (u_e, u_s)$, where $u_s$ is a start UI of the current segment and $u_e$ is the current UI labeled as end. If such back-edge exists then we ``close" the current segment. Otherwise, this UI is added to the current segment and we keep scanning the log. This process stops when the end of the log is reached.

\begin{algorithm}[btp]
{
    \scriptsize
	\Input{Normalised UI log $\bar{\Sigma}$, Back-edges Set $B$}
	\Output{Segments List $\Psi$}
	\BlankLine
	
	Set $\Psi \leftarrow \varnothing$\;
	Set $T \leftarrow$ getTargets($B$)\;\label{alg:targets4}
	Set $S \leftarrow$ getSources($B$)\;\label{alg:sources4}
	Boolean WithinSegment $ \leftarrow$ false\;
	Normalised UI $u_0 \leftarrow$ null\;
	List $s \leftarrow$ null\;
	
	\BlankLine
	\For{$i \in \left[ 1, \mathit{size}\left( \nuilog \right) \right]$}{\label{alg:uilogscan4}
		Normalised UI $\nui \leftarrow$ getUI($\nuilog$, $i$)\;
		\eIf{$\nui \in T$}{\label{alg:segstart4}
			\eIf{WithinSegment $ = false$}{\label{alg:notinsegment4}
				$s \leftarrow$ new List\;\label{alg:startsegment41}
				append $\nui$ to $s$\;
				$u_0 \leftarrow \nui$\;
				WithinSegment $ \leftarrow$ true\;\label{alg:startsegment42}
			}{
				append $\nui$ to $s$\;
			}
 		}{\label{alg:nostart4}
 			\If{WithinSegment $= true$}{\label{alg:insegment4}
				append $\nui$ to $s$\;
				\If{$\nui \in S \wedge (\nui, u_0) \in B$}{\label{alg:startendmatching4}
					add $s$ to $\Psi$\;\label{alg:segmentcomplete4}			
					%$\mathit{Seg} \leftarrow \emptyset$\;
					WithinSegment $ \leftarrow false$\;
				}
			}	
		}
	}
	\BlankLine
	\Return $\Psi$\;
	
	\caption{Identify Segments}\label{alg:segIdentification}
}
\end{algorithm} 
At this point, we collected all the back-edges of the CFG, and we can leverage this information to start segmenting the UI log.
We do so via Algorithm~\ref{alg:segIdentification}.
First, we retrieve all the targets and sources of all the back-edges in the CFG and collect their corresponding UIs (lines~\ref{alg:targets4} and~\ref{alg:sources4}).
Each UI that is the target of a back-edge is an eligible segment starting point (hereinafter, \emph{segment-start UI}),
since a back-edge conceptually captures the end of a task execution, therefore its target represents the first UI of the next task execution.
While, following the same reasoning, each UI that is source of a back-edge is an eligible segment ending point (hereinafter, \emph{segment-end UI}).
%, since it conceptually captures the end of a task execution.
Then, we sequentially scan all the UIs in the UI log (line~\ref{alg:uilogscan4}).
When we encounter a segment-start UI (line~\ref{alg:segstart4}), and we are not already within a segment (see line~\ref{alg:notinsegment4}),
we create a new segment ($s$, a list of UIs), we append the segment-start UI ($\bar{u}$),
and we store it in order to match it with the correct segment-end UI (line~\ref{alg:startsegment41} to~\ref{alg:startsegment42}).
Our strategy to detect segments in the UI log is driven by the following underlying assumption: 
a specific segment-end UI will be followed by the same segment-start UI, so that we can match segment-end and segment-start UIs exploiting back-edge's sources and targets (respectively).
If the UI is not a segment-start (line~\ref{alg:nostart4}), we check if we are within a segment (line~\ref{alg:insegment4}) and,
if not, we discard the UI, assuming it is noise since it fell between the previous segment-end UI and the next segment-start UI.
Otherwise, we append the UI to the current segment and we check if the UI is a segment-end matching the current segment-start UI (line~\ref{alg:startendmatching4}).
If that is the case, we reached the end of the segment and we add it to the set of segments (line~\ref{alg:segmentcomplete4}),
otherwise, we continue reading the segment.

Table~\ref{tab:segments} shows the segment-start and segment-end UIs, highlighted respectively in green and red,
which also ideally delimits the two segments within the normalised UI log.
\begingroup
\renewcommand{\arraystretch}{1.6}
\begin{table}[hbtp]
	\centering
	\scalebox{0.65}{
    \begin{tabular}{c|c|c|l|l}
    & \textbf{UI}
    & \textbf{UI}
    & \multicolumn{2}{|c}{\textbf{UI Parameters and Values}}
    \\\cline{4-5}

    & \textbf{Timestamp}
    & \textbf{Type}
    & \textbf{\textbf{$P_1$: Application}}
    & \textbf{\textbf{$P_2$: Element Label}}\\\hline

    \rowcolor{green}
    1 & 2019-03-03T19:02:18 & Click button &Web & New Record \\\hline
    2 &  2019-03-03T19:02:23 & Edit field &  Web & Full Name \\\hline
    3 & 2019-03-03T19:02:27 & Edit field & Web & Date \\\hline
    4 & 2019-03-03T19:02:39 & Edit field & Web & Phone \\\hline
    \rowcolor{red}
    5 &  2019-03-03T19:02:47 & Click button & Web & Submit \\\hline
    \rowcolor{green}
    6 & 2019-03-03T19:02:58 & Click button & Web & New Record \\\hline
    7 & 2019-03-03T19:03:13 & Edit field & Web & Date \\\hline
    8 & 2019-03-03T19:03:24 & Edit field & Web & Phone \\\hline
    9 & 2019-03-03T19:03:43 & Edit field & Web & Full Name \\\hline
    \rowcolor{red}
    10 & 2019-03-03T19:04:10 & Click button & Web & Submit \\\hline
    \end{tabular}
    }
  	\caption{Segments identification}\label{tab:segments}
  	\vspace{-2mm}
\end{table}
\endgroup

\subsection{Candidate routines identification}
\label{sec:routineIdentification}

%The last two steps of our approach are i) the identification and ii) the selection of the candidate routines.

The candidate routines identification step is based on the CloFast sequence mining algorithm \cite{fumarola2016clofast}. To embed CloFast in our approach, we have to define the structure of the sequential patterns we want to identify.
In this paper, we define a \emph{sequential pattern} within a UI log as a sequence of normalized UIs occurring always in the same order in different segments, yet allowing gaps between the UIs belonging to the pattern.
For example, if we consider the following three segments:
$\langle u_1, u_y, u_2, u_3 \rangle$,
$\langle u_1, u_2, u_x, u_3 \rangle$,
and $\langle u_1, u_x, u_2, u_3 \rangle$;
they all contain the same sequential pattern that is $\langle u_1, u_2, u_3 \rangle$.
%We also recall the notion of support, frequent, and closed sequential pattern.
%Given a sequential pattern:
%its \emph{support} is the ratio of its occurrences and the total number of segments;
%its \emph{frequency}
Furthermore, we define the \emph{support} of a sequential pattern as the ratio of its occurrences and the total number of segments,
and we refer to \emph{closed} pattern and \emph{frequent} pattern (relatively to an input threshold) as they are known in the literature.
By applying CloFast to the set of the UI log segments, we discover all the \emph{closed} sequential patterns.
%Then, we rank the discovered patterns by one of the following criteria: frequency, length, coverage, or cohesion;
%the implementation of our approach allows the user to choose a specific ranking criterion.

Some of these patterns may be \emph{overlapping},
which (in this context) means they were discovered from the same portion of a segment and share some UIs.
An example of overlapping patterns is the following, given three segments:
$\langle u_1, u_y, u_2, u_3, u_x, u_4 \rangle$,
$\langle u_1, u_y, u_2, u_x, u_3, u_4 \rangle$,
and $\langle u_1, u_x, u_2, u_3, u_4 \rangle$;
$\langle u_1, u_2, u_3, u_4 \rangle$ and $\langle u_1, u_x, u_4 \rangle$ are sequential patterns,
but they overlap due to the shared UIs ($u_1$ and $u_4$).
In practise, each UI belongs to only one routine,
therefore, we are interested in discovering only non-overlapping patterns.
For this purpose, we implemented an optimization that we use on top of CloFast.
Given the set of patterns discovered by CloFast, we rank them by a pattern quality criterion (e.g. length, frequency),
and we select the best pattern (i.e. the top rank one).
Then, all its occurrences are removed from the segments, and
we search again for frequent patterns performing the same procedure until there are no frequent patterns left.

In our approach, we integrated four pattern quality criteria to select the candidate routines, they are:
pattern frequency, pattern length, pattern coverage, and pattern cohesion score~\cite{DBLP:conf/iui/DevL17}.
The pattern frequency considers how many times the pattern was observed in different segments.
%Naturally, the most frequent patterns are good candidate routines because their automation may save a significant amount of time.
%On the other hand, longer patterns may be preferable because they describe more behavior.
The pattern length considers the length of the patterns.
The pattern coverage considers the percentage of the log that is covered by all the pattern occurrences.
The pattern cohesion score considers the level of adjacency of the elements inside a pattern,
it is calculated as the difference between the pattern length and the median number of gaps between its elements.
In other words, cohesion prioritizes the patterns whose UIs appear consecutively without (or with few) gaps while taking into account also the pattern length.
%In practice, routines are usually executed in one go without interuptions.
In the next section, we compare these ranking criteria and discuss the benefits of using one or another.

\section{Evaluation}\label{sec:eval}

\urldef{\footurla}\url{https://github.com/volodymyrLeno/RPM_Segmentator}
\urldef{\footurlb}\url{https://doi.org/10.6084/m9.figshare.12543587}

We implemented our approach as an open-source Java command-line application.\footnote{Available at https://github.com/volodymyrLeno/RPM\_Segmentator}
Our goal is threefold. First, we assess to what extent our approach can rediscover routines that are known to be recorded in the input UI logs. Second, we analyze how the use of different candidate routine selection criteria such as frequency and cohesion impact on the quality of the discovered routines. Last, we assess the efficiency and effectiveness of our approach when applied to real-life UI logs.
Accordingly, we define the following research questions:
\begin{itemize}
%\small
\item \textbf{RQ1.} Does the approach rediscover routines that are known to exist in a UI log?
\item \textbf{RQ2.} How do the candidate routine selection criteria affect the quality of the discovered routines?
\item \textbf{RQ3.} Is the approach applicable in real-life settings, in terms of both efficiency and effectiveness?
%scalability and discovering routines that can be recognized by the user?  
\end{itemize}

\subsection{Datasets}

To answer our research questions, we rely on a dataset of thirteen UI logs,
which can be divided into three subgroups: artificial logs, real-life logs recorded in a supervised environment, and real-life logs recorded in an unsupervised environment.\footnote{The real-life logs were recorded with the Action Logger tool~\cite{lenoDemo2019}. All the logs are available at \footurlb}
Table~\ref{table:datasets} shows the logs characteristics. 

\begin{table}[tbh]
%\vspace{-3mm}
\centering
\scalebox{0.85}{
\begin{tabular}{l|c|c|c|c}
\hline
\textbf{UI Log} & \textbf{\# Routine} & \textbf{\# Task}  & \textbf{\# Actions} & \textbf{\# Actions per} \\ 
& \textbf{Variants} & \textbf{Traces}  &  & \textbf{trace (Avg.)} \\ \hline
CPN1 & 1 & 100 & 1400 & 14.000 \\ 
CPN2 & 3 & 1000 & 14804 & 14.804 \\ 
CPN3 & 7 & 1000 & 14583 & 14.583 \\ 
CPN4 & 4 & 100 & 1400 & 14.000 \\ 
CPN5 & 36 & 1000 & 8775 & 8.775 \\ 
CPN6 & 2 & 1000 & 9998 & 9.998 \\ 
CPN7 & 14 & 1500 & 14950 & 9.967 \\ 
CPN8 & 15 & 1500 & 17582 & 11.721 \\ 
CPN9 & 38 & 2000 & 28358 & 14.179 \\ 
Student Records (\studentRecord) & 2 & 50 & 1539 & 30.780 \\
Reimbursement (\reimbursement) & 1 & 50 & 3114 & 62.280 \\ 
Scholarships 1  (\scholarshipA) & & - & 693 &  \\ 
Scholarships 2  (\scholarshipB) & & - & 509 & \\ \hline
\end{tabular}
}
\caption{UI logs characteristics.}
\label{table:datasets}
%\vspace{-4mm}
\end{table}

The artificial logs (CPN1--CPN9) were generated from Colored Petri Nets (CPNs) in \cite{bosco2019}.
The CPNs have increasing complexity, from low (CPN1) to high (CPN9). 
These logs are originally noise-free and segmented. We removed the segment identifiers to produce unsegmented logs. 
  
The \emph{Student Records} (\studentRecord) and \emph{Reimbursement} (\reimbursement) logs record the simulation of real-life scenarios.
The \studentRecord \ log simulates the task of transferring students' data from a spreadsheet to a Web form.
The \reimbursement \ log simulates the task of filling reimbursement requests with data provided by a claimant.  
Each log contains fifty recordings of the corresponding task executed by one of the authors, who followed strict guidelines on how to perform the task. 
These logs contain little noise, which only accounts for user mistakes,
such as filling the form with an incorrect value and performing additional actions to fix the mistake.
For both logs, we know how the underlying task was executed, and we treat such information as ground truth when evaluating our approach. 
%All the present noise is a result of user's mistakes during the execution of the task (e.g., filling in a form with the wrong value). 
Additionally, we created four more logs (\srx , \sry , \srz , \srk) by combining {\studentRecord} and {\reimbursement}. {\srx} and {\sry} capture the scenario where the user first completes all the instances of one task and then moves to the other task.
These logs were generated by concatenating {\studentRecord} and {\reimbursement}. 
{\srz} and {\srk} capture the scenario where the user is working simultaneously on two tasks.
To simulate such behavior, we interleaved the segments of \studentRecord \ with those of \reimbursement. %,i.e.\ the first segment of \studentRecord \ is followed by the first segment of \reimbursement and so on.
%Note, that we used segmented versions of \studentRecord \ and \reimbursement \ in the latter case.

Finally, the \emph{Scholarships} logs (\scholarshipA \ and \scholarshipB) were recorded by two employees of the University of Melbourne who performed the same task. 
The logs record the task of processing scholarship applications for international and domestic students.
The task mainly consists of students data manipulation with transfers between spreadsheets and Web pages.   
Compared to the other logs, we have no a-priori knowledge of how to perform the task in the \emph{Scholarships} logs (no ground truth).
%In other words, we do not have a crisp ground truth, which means that we will have to perform a qualitative evaluation of obtained results. 
Also, when recording the UI logs, the University employees were not instructed to perform their task in a specific manner,
i.e.\ they were left free to perform the task as they would normally do when unrecorded. 

\subsection{Setup}

To answer RQ1 and RQ2, we analyzed the quality of the segmentation and that of the discovered routines, using the first 15 logs described above (CPN1 to CPN9, \studentRecord, \reimbursement, \srx, \sry, \srz, \srk), against the four candidate routine selection criteria in Section~\ref{sec:routineIdentification}, i.e.\ frequency, length, coverage and cohesion. 
%We discuss the results of this part of our experiment from two perspectives: the quality of the segmentation, and the quality of the discovered routines. 
To assess the quality of the segmentation, we use the \emph{normalized} Levenshtein Edit Distance (LED),
%which quantifies how similar two strings are by counting the number of insertions and removals to convert one string into another. 
%The normalized edit distance can be calculated as follows:
%
% \begin{equation}
%	d(x, y) = \frac{editDistance(x, y)}{maxEditDistance(x, y)}
%\end{equation}
%
%The maximal edit distance between strings equals to the length of the longer string. 
%In our context, 
where a segment and its normalized UIs represent the string and its characters, respectively. 
%Two UIs are considered to be equal if they have the same type, and their contexts consist of the same attribute-value pairs.  
Precisely, for each discovered segment, we collect all the ground truth segments that have at least one shared UI with the discovered segment, calculate the LED between the discovered segment and the ground truth segments and assign the minimum LED to the discovered segment as its quality score. Finally, we assess the overall quality of the segmentation as the average of the LEDs of each discovered segment. 

%%%%%% this line GOT TO BE INTEGRATED PROPERLY
%%%%%% We also report on the total number of discovered patterns and their average length, as well as the total coverage.

%The quality of discovered routines is assessed with respect to their length, coverage and their similarity to the ground truth routines, measured with the Jaccard Coefficient (JC).

%JC captures the level of similarity between discovered and ground truth routines in a less strict manner compared to LED. 
The quality of the discovered routines is measured with the Jaccard Coefficient (JC), 
which captures the level of similarity between discovered and ground truth routines in a less strict manner compared to LED. 
In fact, the JC does not penalize the order of the UIs in a routine.
%By contrast to LED, the JC does not penalise the order of the UIs in a routine.
This follows from the assumption that a routine could be executed performing some actions in different order,
and the ordering should not be penalized. 
The JC between two routines is the ratio $\frac{n}{m}$,
%where $n$ is the number of UIs that appear in both the ground truth and the discovered routine, 
where $n$ is the number of UIs that are contained in both routines, 
while $m$ is the total number of UIs in each of the two routines (i.e.\ the sum of the lengths of the two routines).
%$n_{10}$ and $n_{01}$ is the total amount of actions present only in the ground truth or the discovered routine correspondingly. 
Given the set of discovered routines and the set of ground truth routines,
for each discovered routine, we compute its JC with all the ground truth routines and assign the maximum JC to the discovered routine as its quality score.
Finally, we assess the overall quality of the discovered routines as the average of the JC of each discovered routine. As the ground truth, we used the segments of the artificial logs and the guidelines given to the author who performed the tasks in SR and RT. 

However, we cannot rely on the JC alone to assess the quality of the discovered routines,
as this measure does not consider the routines we may have missed in the discovery.
%as it shows only the quality of routines that were discovered.
%On the other hand, it is also important to consider the routines that could be but were not discovered.
Thus, we also measure the total coverage to quantify how much log behavior is captured by the discovered routines.
We would like to reach high coverage with as few routines as possible. Thus, we prioritize long routines over short ones by measuring the average routine length alongside coverage. 

%The quality of discovered routine candidates is measured in the terms of precision, recall, accuracy and F-score given the ground truth.
%For the artificial logs we use the underlying models that were used for logs generation.
%For the logs recorded in controlled settings the ground truth is also known and it corresponds to the sequence of instructions that was given to the user in order to guide him in execution of the task.  
 
%As a part of RQ1 we also evaluate the accuracy of segmentation and study how diversity affects resulting segments and execution time.
%This experiment is conducted on \emph{Student Record} and \emph{Reimbursement} logs.

To answer RQ3, we tested our approach on the \scholarshipA \ and \scholarshipB \ logs and qualitatively assessed the results
with the help of the employees who performed the task. Specifically, we asked them to compare the rediscovered routines and the actions (i.e.\ UIs) they performed while recording. 
%Precisely, we contact the employee of the university who performed the task captured by these logs and ask her whether she is able to recognize discovered patterns and how accurate are they. 

All experiments were conducted on a Windows 10 laptop with an Intel Core i5-5200U CPU 2.20 GHz and 16GB RAM.
%Finally, we present execution time (in seconds) to evaluate computational efficiency.

\subsection{Results}

Table~\ref{tab:segmentationResults} shows the results of the segmentation. 
As we can see, the LED for all the CPN logs is 0.0, highlighting that all the segments were discovered correctly. 
On the other hand, the segments discovered from the \studentRecord, \reimbursement, \srx, \sry, \srz, and {\srk} logs slightly differ from the original ones.
The main difference between the CPN logs and those recorded in a controlled environment is that the former contain routines having always the same starting UI,
while the latter contain routines with several different starting UIs.%\footnote{We discuss this limitation in the next subsection.}
%Accordingly, the last four logs obtained by contatenation of \studentRecord \ and \reimbursement \ suffered from the same limitation.  
%For example, in the \emph{student records} log the first recorded routine starts with the user filling in the form by copying a value from the spreadsheet. 
%The second instance of this task, however, starts with the user clicking on the link to add a new entry and then he repeats the same sequence of actions as in the first instance. 
%In the case of recorded routines with different starting UIs,
%the one that had been observed earlier in the log may dominate all the other starts, and this will affect the segmentation. 
%For all the logs, the correct number of segments was discovered within a reasonable execution time ($<$ 3.5s).

We identified the correct number of segments from all the logs except \srz \ and \srk, 
where we could not discern the ending UI of the routine belonging to one task and the starting UI of the routine belonging to the other task,
consequently merging the two routines and discovering only half of the total number of segments (50 out of 100).
From the table we can also see that the time performance of our approach is reasonable, with maximum execution time of 3.6 seconds. 

\begin{table}[htb]
%\vspace{-3mm}
\centering
\scalebox{0.85}{
\begin{tabular}{l|c|c|c|c}
\hline
\textbf{UI Log} & \textbf{\# Original} & \textbf{\# Discovered} & \textbf{LED} & \textbf{Exec.} \\ 
 & \textbf{Segments} & \textbf{Segments} & \textbf{(avg)} & \textbf{Time} \\ \hline
CPN1 & 100 & 100 & 0.000 & 0.571 \\ 
CPN2 & 1000 & 1000 & 0.000 & 1.705 \\ 
CPN3 & 1000 & 1000 & 0.000 & 0.835 \\ 
CPN4 & 100 & 100  & 0.000 & 0.461 \\ 
CPN5 & 1000 & 1000 & 0.000 & 1.025 \\ 
CPN6 & 1000 & 1000  & 0.000 & 0.707 \\ 
CPN7 & 1500 & 1500  & 0.000 & 1.566 \\ 
CPN8 & 1500 & 1500  & 0.000 & 1.596 \\ 
CPN9 & 2000 & 2000 & 0.000 & 3.649\\ 
\studentRecord & 50 & 50  & 0.059 &  0.714 \\
\reimbursement & 50 & 50 & 0.095 & 1.662 \\ 
\srx & 100 & 100  & 0.078 &  2.424 \\
\sry & 100 & 100 & 0.078 &  2.221 \\
\srz & 100 & 50 & 0.331 &  2.296 \\
\srk & 100 & 50 & 0.331 &  2.536 \\
\hline
\end{tabular}
}
\caption{Segmentation results.}
\label{tab:segmentationResults}
\vspace{-4mm}
\end{table}
\begin{table}[tbh]
 \centering
\scalebox{0.85}{
 \begin{tabular}{c|l|c|c|c|c|c}
 \hline
 \textbf{UI Logs} & \textbf{Selection} & \textbf{\# Discovered} & \textbf{Routine} &
 \textbf{Total} & \textbf{JC} & \textbf{Exec.}\\
& \textbf{Criterion} & \textbf{Routines} & \textbf{Length} & \textbf{Coverage} & & \textbf{Time} \\
 \hline
 \parbox[t]{8mm}{\multirow{4}{*}{{CPN1}}} & Frequency & 1 & 14.00 & 1.00 & 1.000  & 2.643 \\
 & Length & 1 & 14.00 & 1.00 & 1.000 & 1.553 \\
 & Coverage & 1 & 14.00 & 1.00 &  1.000 & 3.702 \\
 & Cohesion & 1 & 14.00 & 1.00 &  1.000 & \textbf{1.530}\\
 \hline
\parbox[t]{8mm}{\multirow{4}{*}{{CPN2}}} & Frequency & 3 & 6.33 & \textbf{0.99} & 0.452 & 3.908 \\
 & Length & 2 & \textbf{14.50} & 0.95 &  \textbf{1.000} & 4.789\\
 & Coverage & 2 & 14.00 & \textbf{0.99} & 0.964 & \textbf{3.166}\\
 & Cohesion & 2 & \textbf{14.50} & 0.95 & \textbf{1.000} & 3.730\\
\hline
\parbox[t]{8mm}{\multirow{4}{*}{{CPN3}}} & Frequency & 4 & 5.75 & 0.95 & 0.511 & 4.682 \\
 & Length & 3 & \textbf{14.33} & 0.93 & \textbf{1.000} & 4.324\\
 & Coverage & 3 & 9.67 & \textbf{0.96} & 0.833 & \textbf{3.940} \\
 & Cohesion & 3 & \textbf{14.33} & 0.93 & \textbf{1.000} & 6.237 \\
\hline
\parbox[t]{8mm}{\multirow{4}{*}{{CPN4}}} & Frequency & 1 &  12.00 & 0.86 & 0.857 & 3.452 \\
 & Length & 2 & \textbf{14.00} & \textbf{1.00} & \textbf{1.000} & \textbf{2.005}\\
 & Coverage & 1 & 13.00 & 0.93 & 0.929 & 3.351\\
 & Cohesion & 2 & \textbf{14.00} & \textbf{1.00} & \textbf{1.000} & 3.655\\
\hline
\parbox[t]{8mm}{\multirow{4}{*}{{CPN5}}} & Frequency & 6 & 1.67 & \textbf{0.86} & 0.206 & \textbf{6.418}\\
 & Length & 7 & 7.29 & 0.83 & 0.849 & 9.715\\
 & Coverage & 4 & 3.75 & 0.80 & 0.462 & 6.587\\
 & Cohesion & 8 & \textbf{7.5} & \textbf{0.86} & \textbf{0.910} & 18.206\\
\hline
\parbox[t]{8mm}{\multirow{4}{*}{{CPN6}}} & Frequency & 3 & 4.67 & 1.00 & 0.485 & 4.250\\
 & Length & 2 & \textbf{10.00} & 1.00 & \textbf{1.000} & 2.924\\
 & Coverage & 3 & 4.67 & 1.00 & 0.485 & \textbf{2.483}\\
 & Cohesion & 2 & \textbf{10.00} & 1.00 & \textbf{1.000} & 4.678\\
\hline
\parbox[t]{8mm}{\multirow{4}{*}{{CPN7}}} & Frequency & 7 & 2.43 & 0.91 & 0.257 & 10.118\\
 & Length & 7 & \textbf{9.57} & 0.88 & \textbf{0.986} & 8.957\\
 & Coverage & 6 & 3.67 & 0.91 & 0.385 & \textbf{7.203}\\
 & Cohesion & 7 & 9.43 & \textbf{0.93} & 0.971 & 11.983\\
\hline
\parbox[t]{8mm}{\multirow{4}{*}{{CPN8}}} & Frequency & 5 & 4.20 & 0.75 & 0.337 & 11.801\\
 & Length & 6 & \textbf{10.67} & \textbf{0.91} & \textbf{0.967} & 9.070\\
 & Coverage & 5 & 7.60 & 0.89 & 0.618 & \textbf{7.354}\\
 & Cohesion & 5 & \textbf{10.67} & \textbf{0.91} & \textbf{0.967} & 11.250\\
\hline
\parbox[t]{8mm}{\multirow{4}{*}{{CPN9}}} & Frequency & 5 & 5.20 & 0.82 & 0.401 & 13.784\\
 & Length & 6 & \textbf{14.67} & \textbf{0.95} & \textbf{1.000} & \textbf{8.265}\\
 & Coverage & 5 & 6.60 & 0.88 & 0.511 & 8.603\\
 & Cohesion & 6 & \textbf{14.67} & \textbf{0.95} & \textbf{1.000} & 13.943\\
\hline
\parbox[t]{8mm}{\multirow{4}{*}{{\studentRecord}}} & Frequency & 3 & 10.00 & 0.96 & 0.356 & 3.883\\
 & Length & 3 & \textbf{28.33} & \textbf{0.98} & \textbf{0.942} & \textbf{2.592}\\
 & Coverage & 2 & 15.50 & 0.96 & 0.532 & 2.635\\
 & Cohesion & 3 & \textbf{28.33} & \textbf{0.98} & \textbf{0.942} & 3.252\\
\hline
\parbox[t]{8mm}{\multirow{4}{*}{{\reimbursement}}} & Frequency & 3 & 18.67 & 0.90 & 0.290 & \textbf{4.63}\\
 & Length & 3 & \textbf{56.33} & \textbf{0.96} & \textbf{0.829} & 5.215\\
 & Coverage & 2 & 30.50 & 0.45 & 0.446 & 4.709\\
 & Cohesion & 3 & \textbf{56.33} & \textbf{0.96} & \textbf{0.829} & 6.585\\
\hline

\parbox[t]{8mm}{\multirow{4}{*}{{\srx}}} & Frequency & 5 & 16.80 & 0.90 & 0.374 & 13.826\\
 & Length & 4 & \textbf{45.25} & \textbf{0.91} & \textbf{0.929} & \textbf{7.362}\\
 & Coverage & 2 & 42.50 & 0.86 & 0.921 & 8.177\\
 & Cohesion & 4 & \textbf{45.25} & \textbf{0.91} & \textbf{0.929} & 10.728\\
\hline
\parbox[t]{8mm}{\multirow{4}{*}{{\sry}}} & Frequency & 5 & 16.80 & 0.90 & 0.374 & 12.176\\
 & Length & 4 & \textbf{45.25} & \textbf{0.91} & \textbf{0.929} & 12.477\\
 & Coverage & 2 & 42.50 & 0.86 & 0.921 & \textbf{9.085}\\
 & Cohesion & 4 & \textbf{45.25} & \textbf{0.91} & \textbf{0.929} & 14.675\\
\hline
\parbox[t]{8mm}{\multirow{4}{*}{{\srz}}} & Frequency & 3 & 28.00 & 0.90 & 0.313 & \textbf{13.905}\\
 & Length & 5 & \textbf{86.40} & \textbf{0.96} & \textbf{0.580} & 28.259\\
 & Coverage & 3 & 55.00 & 0.95 & 0.391 & 14.894\\
 & Cohesion & 5 & \textbf{86.40} & \textbf{0.96} & \textbf{0.580} & 37.277\\
\hline
\parbox[t]{8mm}{\multirow{4}{*}{{\srk}}} & Frequency & 3 & 28.00 & 0.90 & 0.313 & 12.428\\
 & Length & 4 & \textbf{89.50} & 0.90 & \textbf{0.600} & 23.903\\
 & Coverage & 3 & 55.00 & \textbf{0.95} & 0.391 & \textbf{10.838}\\
 & Cohesion & 4 & \textbf{89.50} & 0.90 & \textbf{0.600} & 38.657\\
\hline

 \end{tabular}
}
\caption{Quality of the discovered routines. \label{tab:rediscovery}}
\end{table}

Table~\ref{tab:rediscovery} shows the quality of the discovered routines for each selection criterion,
when setting 0.1 as minimum support threshold of CloFast. 
The results highlight that, overall, the routines with the highest JC and the longest length are those discovered using cohesion as selection criterion, followed closely by those discovered using length as the criterion.
Even though using these criteria we do not always achieve the highest coverage,
the coverage scores are very high, above 0.90 for all the logs except CPN5.
% frequency and coverage selection metrics performed worse than length and cohesion.
%score the highest JC, and identified the longest routines. 
%Compared to other selections metrics, cohesion and length discovered the longest routines with the highest 
%As predictable, also when we choose Both length and cohesion returned longer patterns with a relatively high coverage. 
%The only exception is the CPN5 log. EIGHT
%Using frequency as selection metric allowed us to discover more routines although shorter.% compared to those discovered using cohesion or length selection metrics. 
%Considering that these patterns score a high total coverage, we conclude that they represent different pieces of routine. 
%While using coverage as selection metric we performed better than the frequency-based one in terms of the quality of the patterns,
%and it is also characterized by the best computational efficiency. 

Following these results, we decided to use cohesion as the selection criterion to discover the routines from the \emph{scholarships} logs.
%For RQ3 we selected cohesion-based selection strategy and the mimimum support threshold was set to 0.1. 
From the \scholarshipA\ log we discovered five routines variants.
The first routine variant consists in manually adding graduate research student applications to the student record in the information system of the university. 
The application is then assessed, and the student is notified of the outcome. 
The second routine variant consists in lodging a ticket to verify possible duplicate applications. 
When a new application is entered in the information system and its data matches an existing application,
the new application is temporarily put on hold, and the employee fills in and lodges a ticket to investigate the duplicate. 
The remaining three routine variants represent exceptional cases, where the employee executed either the first or the second variant in a different manner (i.e.\ by altering the order of the actions or overlapping routines executions). 
To assess the results, we showed the discovered routine variant to the employee of the University of Melbourne
who recorded the \scholarshipA\ log, and they confirmed that the discovered routines correctly capture their task executions.
Also, they confirmed that the last three routine variants are alternative executions of the first routine variant.\footnote{Detailed results at \footurlb}
 
%involved in the scholarship management process confirmed that these two routines correspond to the processes they follow.

While the results from the \scholarshipA \ log were positive,
our approach could not discover any correct routine from the \scholarshipB \ log.
By analyzing the results, %with the help of the employee who recorded the log,
%we found out that the employee would systematically overlap routines executions (e.g. starting the next routine before ending the previous).
%Such a behavior is extremely complex to untangle aposteriori, therefore, we conclude that .....
%in order to apply our approach in a real-life scenario.
we found out that the employee worked with multiple worksheets at the same time, frequently switching between them for visualization purposes only. Such behavior recorded in the log negatively affects the construction of the CFG and its domination tree, 
ultimately leading to the discovery of incorrect segments and routines. 
This also had an impact on the execution time, indeed, 
while it took only 41.7 seconds to discover the routines from the \scholarshipA \ log,
it took 426.3 seconds to discover the routines from the \scholarshipB \ log. 

%Such a behavior highlighted another limitation of our approach, 
%because it contained more complex behavior than the one observed in the \emph{scholarships-1} log.
%There, the worker processes student entries from different spreadsheets and continuously switches between them. 
%This affects the control-flow graph, and our technique fails to identify the back-edges. 

\subsection{Limitations}

%Our proposed approach has limitations. 
Our approach relies on  information recorded in the log to identify segments and discover routines. 
Thus, its effectiveness is correlated with data quality. Since the UI log is fine-grained, deviations occurring during the routine execution affect the effectiveness of our approach. In our evaluation, we observed this phenomenon to varying degrees when dealing with real-life logs.
%Thus, it may not work on the logs with poor semantics. 
%Since it heavily relies on the information present in the log, 
%if some information is missing, it can affect the segmentation and the corresponding routines that will be discovered.
In practice, the approach can identify correct routines only if they are observed frequently in the UI log. Recurring noise affects the accuracy of the results (see the \scholarshipB\ log).
%mostly capture the noise-free execution, as long as there are sufficient observations in the input log. 
%Thus, it is possible to miss non-trivial cases. 

The approach discovers multiple variants of the same routine when the UIs of a routine occur in different orders. 
%, rather than clustering all the routine variants into the most frequent one. 
Post-processing the results could be beneficial in order to cluster similar routines.  
%Another limitations are related to the graph structure that is the cornerstone of our approach.
Further, the approach is designed for logs that capture consecutive routine executions. 
In practice, routine instances may sometimes overlap 
(cf.\ scholarshipB\ log).
%(e.g. a user starts working on the next routine before completing the current one). 
%it may happen that a worker will start to work on the next routine before completion of the previous one. 

Finally, while our approach is robust against routine executions with multiple ends,
it is sensitive to multiple starts. Ideally, all  routine executions should start with the same UI,
unless different starts are recorded in batch (e.g.\ first only routines with a start, then routines with another start, etc.).
%This in practice resembles typical business process models.
In general, our approach can handle logs containing multiple different routines, provided that each routine does not share any UIs with other routines, except their start UIs. 

%The structure of the constructed graph depends on the order in which the UIs occur inside the log.
%The UI that had been observed earlier will dominate all the further ones.  
%Since the approach relies on the domination relations to find segments, 
%this implies that the UIs that precede the first occurrence of routine in the log should not belong to any instance of routine or repeat across the log to identify the correct segments. 
 
%We assume that routines recorded in the log are executed consecutively. 
%In the real-life, however, it is common that the worker may start to work on the next routine before completion of the previous one. 
%If a certain action occurs before the actual start of routine and is present inside routine, it will create additional entry points in the corresponding strongly connected component. 
% !TEX root = main.tex
\section{Conclusion and Future Work}\label{sec:conclusion}

This paper presented an approach to automatically identify routines from unsegmented UI logs. The approach starts by decomposing the UI log into segments corresponding to paths within the connected components of a Control-Flow Graph derived from the log. Once the log is segmented, a noise-resilient sequential pattern mining technique is used to extract frequent patterns. The patterns are then ranked according to four quality criteria: frequency, length, coverage, and cohesion. 

The approach has been implemented as an open-source tool and evaluated using synthetic and real-life logs. The evaluation shows that the approach can rediscover routines injected into a synthetic log, and that it discovers relevant routines in real-life logs. The execution times range from seconds to a few dozen seconds even for logs with tens of thousands of interactions.

As future work, we aim to address the limitations discussed above. We plan to add a post-processing step to group multiple routine variants and to discover an aggregated model thereof. This could be achieved by clustering the patterns and merging them into high-level models or by adapting a local process mining technique~\cite{DBLP:journals/topnoc/DalmasTN18}. We plan to design more sophisticated segmentation techniques to better handle mixtures of multiple routines. Finally, the approach identifies routines from a control-flow perspective insofar as it manipulates sequences of interactions, without considering their data payload. We plan to complement the approach with an approach to quantify the automatability of candidate routines based on data attributes.

%We also consider an alternative approach to discovering candidate routines without preliminary segmentation.
%One of the possible improvements would be to devise a technique to automatically identify context attributes, and thus, make the approach more independent. Also, at the moment, the approach requires expert knowledge to correctly identify the segments.  

\medskip\noindent\textbf{Acknowledgments.} Work supported by the European Research Council (PIX project) and by the Australian Research Council (DP180102839).

\bibliographystyle{IEEEtran}
\bibliography{main} 

\end{document}